# Malicious User Experience Design Research for Cybersecurity

Adam Trowbridge
DePaul University
243 S Wabash Ave
Chicago, IL
atrowbr1@cdm.depaul.edu

Filipo Sharevski
DePaul University
243 S Wabash Ave
Chicago, IL
fsharevs@cdm.depaul.edu

Jessica Westbrook
DePaul University
243 S Wabash Ave
Chicago, IL
jwestbro@cdm.depaul.edu

## ABSTRACT
This paper explores the factors and theory behind the user-centered research that is necessary to create a successful game-like prototype, and user experience, for malicious users in a cybersecurity context. We explore what is known about successful addictive design in the fields of video games and gambling to understand the allure of breaking into a system, and the joy of thwarting the security to reach a goal or a reward of data. Based on the malicious user research, game user research, and using the GameFlow framework, we propose a novel malicious user experience design approach.

**Security and privacy → Human and societal aspects of security and privacy →** Usability in security and privacy

## 1. INTRODUCTION
In a paper that outlines a framework for understanding the user experience of game players, or player experience (PX), Nacke and Drachen [1] point out that both the game system and the player are actors exerting influence on the experience of playing the game. The game system, in this case meaning the combination of what the game can do, what rules constrain play, and what affordances are offered to the player, creates the basis for the play experience. The player provides the context for the game, entering data via the game interface and receiving output via audio visual cues, causing changes in state that constitute playing the game.

Applied to cybersecurity, this template reveals the secured system as providing the functionality of a sort of game, blocking some actions via constraints, and potentially offering other opportunities via vulnerabilities. The malicious user provides the context, entering data and receiving feedback, and may cause changes in the state that constitute a hack. This similarity, we argue, may not only appear on the surface in this comparison, and we use the comparison of game user and malicious user as a basis for design research into a system focused on meeting the needs of malicious users (below, we define "malicious user" to include any unauthorized user, regardless of motive, including profit [2]).

In this paper we will explore the factors, and conduct initial user-centered research, necessary to creating a successful game-like prototype, and user experience, for malicious users. The background research requires investigating factors that make a successful game, including what is known about successful addictive design in the fields of video games and gambling. We will review the traits of the user base addicted to those experiences and compare that user base to what is known about malicious users. The comparison of these two groups is intended to move beyond criminology-based theories for malicious user motivation and move our malicious user design project closer to designing experiences to specifically engage malicious users as if they were game users. In this, we are informed by Xu, Hu, and Zhang's study of six computer hackers in China as well as other research focused on explaining the behavior of malicious users [3]. While Xu et. al. combine routine activity theory, social learning theory, and situational action theory to explain the progression of computer use to malicious computer use, no part of that explanation touches upon the motivation provided by the systems themselves, or the activity necessary to hack them. This lack was common across criminology-based discussions on the personality traits of hackers [4]. Missing in most of these studies is a discussion of the hacking activity itself, the allure of breaking into a system, and the joy of thwarting the security to reach a goal or a reward of data.

Flow theory and research seek to understand the "phenomenon of intrinsically motivated, or autotelic, activity: activity rewarding in and of itself (auto self, telos goal), quite apart from its end product or any extrinsic good that might result from the activity" [5]. Flow is a state of complete absorption in an activity, immersion is the experience of the flow state. While Rennie and Shore discuss the application of flow theory to the pleasure of hacking, they primarily approach this model in relationship to increasing proficiency in malicious computer use [6]. The work most aligned with our own research is that of Kevin F. Steinmetz's discussion of hacking as a transgressive craft [7]. In his dissertation, he proposes seeing hacking as "a behavior driven by a desire to learn, grow, and excel," and even points out that that the end product may not be as important as the activity of hacking [7]. He further describes hackers as creative, within the transgressive craft framework, doing things with tools in ways that were not initially intended. He also frames hacking communities as craft guilds, who work alone on individual projects, but come together to share information and improve the craft.

As part of this exploration, we pose what might otherwise be termed "hackers" as a class of users, within a user experience design, or user-centered design, context. The term "malicious" is used as a signifier to differentiate what are commonly referred to as "hackers" from the users that computer software designers consider when designing systems, without excluding hackers from being considered as users of the system. Malicious users are different users than those considered in designing experiences for users of a system and are nearly always discussed as attackers rather than users. However, we argue that they remain a category of user in relationship to every system, especially networked systems. These malicious users experience a very different User Interface (UI), accessing systems via their own tools, via a command line showing responses from the system, or potentially creating their own UI. These users rely on both unintended and intended discoverability, and they are not interested in following the constraints of either the interface or the security.

In looking for a user class that most closely parallels the attraction to a craft-like approach to a user interface, as well as the guild-like system of communication and sharing of knowledge and craft, we posit that hackers may be very similar to game users, particularly game users who may be described as addicted to, or with markers for potential addiction to, games. The similarity is underscored in Steinmetz's description of the activity of hacking: "often hacking is a solitary activity involving long hours in front a computer or other piece of technology, working away at it to accomplish a task or learn more" [7]. His description of the importance of community for learning could equally describe game communities: "in this sense, learning in a group setting accelerates the rate of learning by filling in the gaps in knowledge which might otherwise take a tremendous amount of time to overcome" [7]. Expanding malicious user research to include research on gaming and problem gaming allows us to develop a broader potential profile of malicious users, particularly in regard to immersion and flow states.

Whereas most software is focused on usability in service of productivity, games are designed to provide entertainment [1]. Fields that involve usability tend to describe usability in service of functionality and productivity rather than entertainment and immersion. The field of Human Computer Interaction (HCI), rooted in ergonomics (the study of human efficiency in working environments) is tied to similar concerns in the workplace [8]. There are also fields of amusement and entertainment design, parallel to ergonomics, but, similar to the relationship between serious usability studies and game studies focused on making games more usable (i.e. fun), there is little crossover in studies between the two fields. These factors further underscore the "outsider" status of both malicious users and game users (as well as game designers), in relationship to productivity.

However, there is increasing crossover between productivity-based user experience design and game mechanics in the form of gamification: the integration of game elements and game mechanics in productivity-related software [8]. As an example of game use in cybersecurity, picoCTF (capture-the-flag, CTF) is a security competition with a web-based game interface [9]. The competition uses a capture-the-flag game construct to "encourage greater computer science interest among high school student" [9]. While picoCTF employs a game to introduce cybersecurity to students, it does not gamify the malicious user experience. The picoCTF game was announced publicly as a game, students signed up for the competition, and were thus aware they were playing a game. Students playing picoCTF took on the role of malicious users but were not malicious users. The nature of the game was specifically designed as a visual story game called Toaster Wars, "to appeal to students who might not otherwise be interested in participating in a computer security competition" [9]. Capture-the-flag has been employed as a cybersecurity game since at least 1996 [10]. In a Jeopardy-style CTF games, players attempt to find hidden information (flags) planted in the game space. This "space" may be file systems or network-based services. Players take on the role of malicious users and attempt to exploit vulnerabilities to reach the flags.

While CTF games approach cybersecurity training as play and locates that play inside the usual spaces of hacking (file systems and network-based services), they do not seek to immerse the game users. Players are aware that they are playing a game, acting as attackers and, in some CTF games, as defenders. The picoCTF game adds a narrative, and thus some level of immersion, to cybersecurity activity, but the visual overlay moves it outside the spaces that immerse actual malicious users: file systems and network-based services. While both may inform malicious user design, neither are specifically designed for malicious users.

Some malicious users already approach hacking as a game. Turgeman-Goldschmidt claims that hackers "are offering society new rules for play" [11]. and that the hackers see their activity as fun, even thrilling, and ultimately not serious, despite also believing it to be subversive and a form of resistance to social order. This paper borrows from gamification approaches, to propose malicious user experience design as closer to game design (engagement and immersion) than software design (usability and productivity), or an extension of criminology. Overall, the goal of this research is to provide sufficiently solid user research on which to base a malicious user experience design prototype that can be combined with rapid testing and iteration.

## 2. GAME USERS, ADDICTIVE PLAYERS AND PROBLEM GAMBLERS

Defining the nature of addiction, and the point at which gaming (including gambling) becomes a problem is outside the scope of this paper. Instead, we will focus on factors that lead to intense interaction with both video games and video gambling, including literature that attempts to define and describe addiction. Researchers focused on game-related studies delineate between positive gaming behavior and behavior that becomes problematic. This is often defined at the point that gaming results in negative consequences [12]. However, from the perspective of cybersecurity (and, in most cases, a legal perspective) malicious users are engaged in behavior with negative (potential) consequences as soon as they begin using a system maliciously. From these perspectives, we pull equally from game research into "acceptable" game use and potentially problematic game use in order to conduct user research for malicious user experience design.

### 2.1 General traits and motivations

Kuss and Griffiths concentrated multiple findings into three key personality traits related to internet gaming addiction: "introversion, neuroticism, and impulsivity" [13]. Emond et al. found that problematic gambling was "inversely related to rational thinking and positively related to experiential thinking" [14]. They further state that experiential thinking may inhibit the use of instruction on gambling and probability having an effect on reducing gambling behavior, due to a resistance to rational/analytical thinking. Their study found that "dysfunctional coping, socialization and personal satisfaction serve as risk factors for developing Internet gaming addiction" [14]. In terms of detrimental gambling related conditions, the study found an inverse relationship between rational thinking and detrimental gambling, as well as a positive relationship between experiential thinking and detrimental gambling [14]. The same research also found that gambling behavior was resistant to education about probability (rational thinking).

## 2.2 Social

Loneliness and decreased social competence are experienced by game players who have high scores but also problems with compulsive game use [12]. At the same time, sociability and social elements of games "significantly predicted video game addiction" [12]. Game users with problematic game behaviors were also more likely to play games with a high social component "e.g., sharing tips and strategies, cooperating with other players, etc." [15].

## 3. THE MALICIOUS USER

Research into hacker motivation and personality is limited by the need to seek out people engaged in activity that is most often illegal, as well as socially unacceptable within wider society. Hackers engage in malicious use of systems for multiple reasons, including for financial gain [2], [16]. However, the research we reviewed on malicious users did not specifically investigate financially-motivated malicious users. Their approach, and ours, treats hacking as a specific set of actions and defines hackers as those who engage in that activity.

Pulling from existing studies of hackers that go beyond motivation and profiling from criminological theory, we have a rough outline of malicious users. They are more analytical and rational in their thinking, and have higher confidence in their decision making [17]. They see their actions as a form of entertainment rather than crime [11]. They are more open to change than they are concerned about conserving safety and security [18]. They rely on a social network consisting of different levels of skill ability, and give back to that community when they learn something new [7].

## 3.1 General traits and motivations

In a grounded theory study "designed to examine the social construction of the reality among Israeli computer hackers through the accounts they use to explain their deviant behavior," Orly Turgeman-Goldschmidt conducted 54 interviews with Israeli hackers and found that they were motivated by fun, thrill, and excitement; curiosity and exploration, including voyeurism; power and dominance; economic justice; revenge; and ease of use [11]. Among other places, participants were sourced via advertisements in media, targeted ads at hacker conferences, and security conferences resulting in 51 interviewees that were men, with ages ranging between 14 and 48.5 years old. The research involved "approaching the hacker community to discover relevant categories and the relationships among them" via narrative interview [11]. Based on these interviews, Turgeman-Goldschmidt claims that "hacking is a new form of entertainment based on the play-like quality that characterizes the use of digital technology and is a new form of social activity" [11]. The hackers in his study see themselves as pursuing their malicious computer user goals in line with "values that are praised in today's society: the pursuit of happiness, curiosity, and knowledge and the demonstration of computer virtuosity" [11]. In a review of malicious user motivation, Madarie found that openness to change, rather than safety and security values, correlated with willingness to circumvent system security. However, the study also found that intellectual challenge and curiosity are not related to frequency of hacking, leading to speculation that hackers may be "motivated by what they dislike instead of being motivated by what they value" [18].

In a study on the personality characteristics of illicit computer hackers, Michael Bachmann found that successful hackers have a "strong preference for rational decision-making processes" [17]. This is a significant difference from the thinking style found in problematic game users. This disparity is key to our malicious user experience design research. If immersion, from a flow theory perspective, is important to capturing and holding the attention to gamers, and that immersion is rooted in gamers' experiential personalities, where does that leave us if we want to capture and hold the attention of malicious users, who skew toward rational/analytical thinking rather than experiential? Malicious user experience design approaches will need to be immersive for rational/analytical thinkers.

## 3.2 Social

The malicious users in Turgeman-Goldschmidt's study disputed the popular conception of hackers as loners, preferring lone computer use to human company [11]. Steinmetz sees the sharing of acquired knowledge as vital to the hacker community and suggests that there is social pressure to not only learn and improve, but to share what one has learned to help the entire hacker community to improve. The communities of learning he describes include passive learning via accumulated knowledge available online and active community events designed for knowledge sharing [7].

## 4. FLOW THEORY AND CYBERSECURITY: THE RELEVANT PARTS

Flow theory attempts to provide a model for enjoyable experiences, particularly those that are seriously engaging, or immersive, to the point of losing track of time, or having an altered sense of time, while the activity continues. The theory approaches the elements of enjoyment as universal [19]. Flow experiences consist of eight elements, as follows:

(1) a task that can be completed;
(2) the ability to concentrate on the task;
(3) that concentration is possible because the task has clear goals;
(4) that concentration is possible because the task provides immediate feedback;
(5) the ability to exercise a sense of control over actions;
(6) a deep but effortless involvement that removes awareness of the frustrations of everyday life;
(7) concern for self disappears, but sense of self emerges stronger afterwards; and
(8) the sense of the duration of time is altered.

The combination of these elements causes a sense of deep enjoyment so rewarding that people feel that expending a great deal of energy is worthwhile simply to be able to feel it [15].

## 4.1 Flow theory and gaming

Sweetser and Wyeth adapted flow theory to gaming, creating the following new GameFlow categories: concentration, challenge, player skills, control, clear goals, feedback, immersion, social interaction [19]. Combined, these indicate that a game authored to create flow state should require and allow concentration; the game challenge should match the player's ability; the player must be able to progress in skill to the point of mastering the game; the player should feel a connection between taking action and seeing a result in the game; the goals of each stage of the game should be clear; the game must provide feedback; the player should be immersed, becoming less aware of anything outside the game, losing sense of time; and the game should offer the opportunity for social interaction [19].

In a study of the relationship of appeal and immersion in video games, Georgios Christou highlights the importance of initial appeal in getting a game user to engage with the game, and the importance of immersion in drawing users into continuing to play and return to a game [20]. A study of flow as a predictor of video game addiction identified one factor, out of the nine original flow theory elements, as significant: "heightened levels of a sense of time being altered during play" [15]. To counter the loss of time that is prevalent for game users prone to addiction, the study suggests that games should make players aware of time passing. Like casinos, without clocks and wrapped in mazes of engagement, malicious user experience design should seek to do the opposite, to encourage immersion and loss of temporal awareness.

## 4.2 Flow theory and cybersecurity

Cybersecurity attempts to interrupt flow states. While studies on the relationship of flow state to gaming addiction delineate between the positive nature of a flow state and the point at which desire for the state becomes an addiction, from the perspective of cybersecurity, any flow state in a malicious user is negative (aside from flow achieved from pointing in the wrong direction, or at the wrong target). Rennie and Shore suggested that increased system security could interrupt flow states and progression as a hacker by reducing the initial effectiveness of scripts and tools used by "script kiddies" (i.e. entry-level malicious users that rely on tools built by more advanced malicious users) [6].

Sweetser and Wyeth's GameFlow model posit flow as the singular goal of a game, a self-contained, rewarding experience. The reward is "a sense of discovery, a creative feeling of being transported into a new reality" [19]. We are left with the questions raised in section 3.1 regarding the experientiality of game users and the rationality of malicious users. In *Neurocognitive mechanisms underlying the experience of flow*, Dietrich claims that "a necessary prerequisite to the experience of flow is a state of transient hypofrontality that enables the temporary suppression of the analytical and meta-conscious capacities of the explicit system" [21].

This seems to support our concern regarding whether malicious user experience design can craft a flow state when the potential user base skews toward analytical thinking. However, further in the paper Dietrich discusses the merger of action and awareness and describes this form of immersion as the disappearance of self-consciousness, no worry of failure, a sense of timelessness, and no distractions. From this viewpoint, remembering the self-confidence characteristic described by Bachmann [17], in the case of designing for malicious users, immersion is not experiential immersion but an engagement that supports self-confidence, encourages intensity that results in a loss of the sense of time, and provides no distractions (including the necessity of *opacity*, creating a belief that the system being used maliciously is not misleading or monitored).

## 5. HONEYPOT DESIGN: MALICIOUS USER EXPERIENCE DESIGN?

On the surface, much of the potential value of our research is already covered by the use of honeypots, which "lure a potential attacker by simulating resources having vulnerabilities and observing the behavior of a potential attacker to identify him before a damaging attack takes place" [22]. High-interaction honeypots can be tools to collect intelligence about what motivates malicious users and how they operate while hacking a system [23]. Gupta et al. extend the honeypot construct to include internal privacy controls, using a system they refer to as a "context honeypot". In their proposed system, malicious users are lured using fake data that is similar to the real data a suspected (internal) malicious user is likely to see [24]. They further explore and quantify the concept of "opaqueness" as the aspect of the system that creates the impression that the honeypot is real, the responses are real, and the data they are accessing is real [25]. This is useful background research for the concept of malicious user experience design.

However, honeypots, even "context honeypots" that anticipate potential malicious use and adapt to provide synthetic data similar to the sort of data the malicious user is seeking, as well as "high-interaction" honeypots that include a full experience, remain firmly in the domain of criminology research and standard cybersecurity. Creating a honeypot is analogous to dropping a $20 bill on the ground to see who picks it up. The context honeypot is analogous to putting a camera in the lunchroom at work and stocking the refrigerator with food cartons. The honeypot provides an opportunity for crime, and may cross over to entrapment, inducing people to commit crimes that would not ordinarily commit [25], but even a honeypot that targets a specific user or class of users with specific types of data is a very shallow authored experience. We are interested in crafting a far more targeted immersive experiences, that are specifically user-centered, and crafted toward long-term "playability". Beginning with a honeypot in mind would be far too limiting. Further research into honeypots, both as a starting point for luring and as a source of information about opaqueness, is warranted. Given the level of interaction available via existing honeypot software and approaches, we do not see the results of this research (ultimately, prototype systems) resembling existing honeypots.

## 6. DIRECTIONS: FACTORS IN DESGINING FOR THE MALICIOUS USER

Based on the malicious user research, the game user research, and using the GameFlow framework, we can begin to outline a malicious user experience design approach. GameFlow applied to malicious users sketches a rough outline of the prototype: challenging tasks with clear goals that can be accomplished, and a system that provides immediate feedback. We are interested in crafting targeted immersive experiences that are specifically malicious user-centered and crafted toward long-term "playability".

### 6.1 Deception

Bob Blakley says that a malicious user "should get a really bad experience, that we're designing to be bad" [26]. Blakley suggests the user experience goals for this approach to design are: "confusion, expense, and difficulty" [26]. While considering login feedback, Blakley asks why the system does not respond to a failed attempt with "welcome to your extremely convincing simulated fake account" [26].

The system designed for malicious users' needs to provide a similar experience as Blakley described, but perhaps for different reasons. The system must provide an environment in which the user can concentrate, it must get the user's attention, and hold it. Traditional, low-interaction honeypot design would include a system with seemingly flawed security and what seems like worthwhile data. Malicious user experience design needs to hold the users' attention for longer. Opacity is defined as characteristics of the system that make it believable. The opacity of the design is important to concentration. Any indication that the system has another purpose, or that it is not responding normally, will distract and ultimately drive away the user [24].

In order to avoid arousing suspicion, the malicious user's experience must provide the expected level of Blakley's factors (applied to a different purpose): confusion, expense, and difficulty.

The system must be sufficiently complex and require significant time and effort to use. The design should provide challenges that matches the user's skill level (which might be determined by the vulnerability that allowed the user entry into the system). In a static system, this would mean that the system must be designed for a particular user skill level or offer multiple diverging paths (different vulnerabilities and data payloads). Designing multiple systems may interrupt the opacity of the design, and multiple redesigns will be required for each individual "level".

## 6.2 Adaptive game balancing

We believe that the system may be dynamic and scale to the ability level of the user, as many games do via dynamic game balancing. Andrade et al. found that adaptive game balancing, in which the system adjusts difficulty to the players ability using artificial intelligence, had the highest satisfaction rating in usability tests [27]. The design should offer malicious users the ability to develop their skills. Unlike designing a game, we cannot make up a set of mechanics that the player masters. In order to maintain opacity, and thus concentration, the design must be based on existing, or possibly emerging, security flaws, system configurations, and cybersecurity approaches. These could be adaptive, and a robust system could offer multiple jumping off points that change and adapt to existing users. Users must be rewarded regularly as they master specific parts of the system, either with data that appears real and valuable, or with additional access to more of the system.

The system must respond in such a way to maintain the illusion that the user is in control and the system is appropriately complex. As a simple example, in a command line scenario, the system should name system files appropriately but may create data file names that are attractive to a malicious user. Taken to the level of parody, this would result in files named "passwordhere.txt, breaking the immersion and making the systems goals transparent. As with any user experience design, malicious or not, the system must balance between pushing the agenda of the system's creators and meeting the needs of the user. As Blakley put it, "welcome to your extremely convincing simulated fake account" [26].

## 6.3 Manufactured reality

In order to maintain opacity, each "stage" in the authored experience for malicious users must clearly lead to the next ("clear goals"), via the expected response from the system. This approach would require a delicate balancing act between making the reward (data or access) apparent, but only after the player explores the system further. Immersion must come from all of the above, with the possibility that the combination of increasing access and data rewards begin to tell a story, however rough, that leads the user forward. From the perspective of the rational/analytical thinker, this story must maintain opacity, meaning that the system is believable and the ultimate goal is sufficiently difficult to attain.

In a paper on deception-based defenses, Almeshekah and Spafford describe the goals of these defenses as: "manufacture reality, alter reality, and/or hide reality" [28]. They suggest the creation of "plausible alternatives to reality" in order to deceive attackers and suggest that these realities should be based on "specific biases in how people think" [28]. Immersion, above all, appears to be the user research-based goal that is key to malicious user design. Malicious user design must manufacture realities that are plausible to malicious users with a strong preference for rational thinking. Almeshekah and Spafford's review of categories of cultural biases will be instructive in designing malicious user focused systems that provide plausible realities.

## 6.4 Social interaction

Finally, and secondary to immersion in importance to both game and malicious users, is social interaction. We know that game users and malicious users communicate in small groups, read information published by other users, and congregate online and in person. Designing an adaptive system for a single user without reducing or eliminating opacity seems possible. Doing so for a user base that communicates regularly (and in places not readily available to the designers) is far more difficult. A system cannot adapt to multiple players and, at the same time, seem like a real system with flawed security, if those users compare notes.

However, the research regarding social interaction is helpful in considering vectors for learning whether communication about our system reaches hacker communities, and potentially getting feedback on whether there is discussion about our current system iteration, and whether it is convincing. If it is not, monitoring hacker social interaction could prove valuable in the next iteration. More experimentally, and forward thinking, could a system use Artificial Intelligence (AI) and machine learning to provide more immersive interaction? The complexity of applying AI to social engineering and the challenge of constructing a convincing "chat bot" that passes for human is beyond the scope of our current research. However, some simpler uses of Markov chains to determine the location of phishers is promising in showing how AI communication could, at minimum, be used to draw malicious users toward a system [29].

## 6.5 A system for the malicious user

We expect that the initial prototype(s) will be indistinguishable by some from a honeypot, and especially from a context honeypot. It will likely be limited in scope and need to have a beginning and end, with a specific goal or set of goals the user is guided toward. Further research is needed into variations in malicious user skill level. This would allow us to develop at least a limited scope of the levels of game challenge, e.g. "beginning," "intermediate," and "advanced."

The malicious user design approach does not begin with the goal of collecting data on malicious users any more than Facebook and Twitter begin with the goal of collecting data on users, which is to say that it is the goal but only by way of creating a construct for users to provide data while doing other things. Like those corporate, social media projects, a malicious user design must create an experience on which to base the collection of data. In our approach, data collection is intended for user research and research documentation. We have not identified a goal for the system to complete, other than immersing the malicious user as long as possible, to create and maintain a flow state. In this, our prototype goals are similar to those of designers of video gambling machines. Those systems are intended to keep their users immersed until they run out of money or spend so much money that they can no longer maintain a flow state. Our research is not intended to build either an offensive or defensive tool for cybersecurity. We feel that remaining focused on malicious design earnestly, with the same care that immersive game designers take in crafting games, offers more opportunities for innovation in malicious user design and in cybersecurity.

## 7. DISCUSSION AND LIMITATIONS

The initial research into game users and malicious users in this paper seems promising, in terms of approach, but it is far from concrete. User research in the field, and in practice in industry, is often based on incomplete information coupled with prototyping and testing. Our goal in this paper was to collect and analyze initial user research upon which to build a prototype for malicious users, and in this, we feel we have enough to go on.

## 7.1 Approach limitations

The GameFlow framework provides one possible approach to malicious user design, and one that appears to match our interest in approaching malicious user design in parallel to research based on game users. However, the GameFlow framework is an adaptation of flow theory that has not been developed as a full evaluation tool for games, and the flow theory it was based on is similarly complex and difficult to prove definitively. Our experimental research with a malicious user design prototype may contribute to research on flow as it relates to hacking, and flow within games.

## 7.2 Technical requirements

Adaptive game balancing offers a conceptual research direction that can scale to the level of our prototype implementation. The primary questions driving technical limitations is how deep and long can the flow state of immersion be maintained. As an example scenario, we can imagine an initial prototype made up of a simulated command line (which may include or use an actual command line), files that contain data the user is lead to believe are valuable (via location and name) but which contain data that prods the user to further explore the system. Adaptive game balancing in this simple prototype may adapt the difficulty of vulnerabilities based on the exploit the user employed to access the system. There may be multiple access points representing different systems. At the present time, the scope of designing the content (scenarios) of an initial simple prototype is more complex than designing the technical system to support it.

## 7.3 Potential future research

We intend to apply a quick iterative approach to prototyping this system, although "quick" will be a relative term in relationship to the complexity of the task at hand. The next stage of development, iterations of prototypes based on the research at hand, can occur in parallel to further malicious user research, including social research. We feel we have only scratched the surface in the comparison of malicious users to game users.

There is a distinct opportunity to further explore social engineering as it relates to the malicious user design profile contained in this paper. There is a relatively recent field of game design referred to as "alternate reality" gaming, which mixes interaction and game events that occur in real life and online. These games rely heavily on online social interaction between players that is outside the control of the designers, but that is sometimes influenced by "actors" who participate in the game to invisibly guide the interaction toward specific goals. Developing an alternate reality game requires socially engineering the players. Alternate reality game design includes all of the game factors discussed above, as well as complexly timed events and non-linear storytelling Its potential as a model for malicious user design may add unnecessary complexity to an already complex initial prototype. However, future research may include iterations based on alternate reality game approaches. Additionally, the system we have described may have uses for researchers in addictive games, online games, and alternate reality games. Our system might also be useful to better understand the health aspects of video game addition, or "gaming disorder," which is being added to the World Health Organization's new draft list of diseases, with the classification expected to become official in 2019 [30].

## 7.4 Ethical, implementation, and experimentation limits

As mentioned above, honeypots have been discussed as possibly crossing over into entrapment, inducing people to commit crimes that would not ordinarily commit. People attempting to access a honeypot believe they are, by definition, committing a crime or, at minimum, accessing a system or account without the owner's permission. They will likely attempt to mask their identity in multiple ways. For example, it is not plausible to identify the age of a malicious user visiting a honeypot.

The use of this system will require time and a segment of the population of malicious users conduct their activities for multiple reasons, including finical gain. There are ethical implications in creating a system that distracts users from their intended purpose for as long as possible. Our review of research includes evidence that many malicious users are not adults, and thus our prototype could conceivably attract users who are legally children. These issues are already a consideration in existing games, as well as social media [31]–[33].

Renaud and Warkentin consider risk in information security research as it relates to ethics and, specifically, to Institutional Review Boards (IRB) in the United States [34]. They speculate that a typical IRB would not approve research that puts subjects at risk for encountering malware on their personal device. Given the lack of consent necessary for an opaque malicious user design experience (the user cannot know they are part of an experiment and thus cannot consent), the lack of identifying factors that would allow excluding children from the research (or even knowing whether children were included), and the fact that to participate in the system at all, users must believe they accessing the system without the owner's permission, it seems possible that our prototype will never be implemented as university research. However, Renaud and Warkentin suggest that we find a way forward directly, and that "we should write well-argued motivations for our studies, in terms of benefits to society as a whole; Assuming that such studies would be turned down is perhaps overly naïve" [34].

We believe that our research benefits the society as a whole by offering a divergent approach to researching malicious users, who are a worldwide threat to privacy and security. The research is a combination of an unorthodox, interdisciplinary perspective and approach, and a willingness to put together interdisciplinary user research to move toward a prototype. The next step in this research is to build a prototype using the GameFlow framework, based on "plausible alternatives to reality," and targeted toward the general profile of malicious users contained in this paper.